# A source of high–velocity white dwarfs

Melvyn B. Davies[1], Andrew King[1], and Hans Ritter[2]
[1] *Department Physics & Astronomy, University of Leicester, Leicester LE1 7RH, UK*
[2] *Max–Planck–Institut für Astrophysik, Karl–Schwarzschild-Str. 1, D–85740 Garching, Germany*

28 October 2018

**ABSTRACT**
We investigate whether the recently–observed population of high–velocity white dwarfs can be derived from a population of binaries residing initially within the thin disk of the Galaxy. In particular we consider binaries where the primary is sufficiently massive to explode as a type II supernova. A large fraction of such binaries are broken up when the primary then explodes as a supernova owing to the combined effects of the mass loss from the primary and the kick received by the neutron star on its formation. For binaries where the primary evolves to fill its Roche lobe, mass transfer from the primary leads to the onset of a common envelope phase during which the secondary and the core of the primary spiral together as the envelope is ejected. Such binaries are the progenitors of X-ray binaries if they are not broken up when the primary explodes. For those systems which are broken up, a large number of the secondaries receive kick velocities $\sim 100 - 200$ kms$^{-1}$ and subsequently evolve into white dwarfs. We compute trajectories within the Galactic potential for this population of stars and relate the birthrate of these stars over the entire Galaxy to those seen locally with high velocities relative to the LSR. We show that for a reasonable set of assumptions concerning the Galactic supernova rate and the binary population, our model produces a local number density of high–velocity white dwarfs compatible with that inferred from observations. We therefore propose that a population of white dwarfs originating in the thin disk may make a significant contribution to the observed population of high–velocity white dwarfs.

## 1 INTRODUCTION

Oppenheimer et al (2001) have reported a population of high-velocity wide dwarfs in the solar neighbourhood which they claim to be halo objects. From their observations they infer a local number density of such objects to be $\sim 10^{-4}$ pc$^{-3}$ much larger than previous estimates (Gould, Flynn & Bahcall 1998). Reid et al (2001) however, claim these stars are simply the high–velocity tail of the so-called thick disk population. This result is controversial; for example Hansen (2001) notes that the ages of the white dwarfs appear to have a spread, whereas the thick disk stars have similar ages (around 12 Gyr old).

We investigate whether the observed population can be derived from the binary population within the thin disk. Recently, Koopmans & Blandford (2001) suggested that young white dwarfs could be ejected from the thin disk to the halo, through orbital instabilities in triple–star systems. In this paper, we consider a simpler scenario, where the secondary of a tight binary is ejected when the primary explodes as a supernova, breaking up the binary through the combined effects of mass loss in the supernova explosion and the kick received by the neutron star. To produce the high–velocity objects seen by Oppenheimer et al, one must impart kicks in the range 100 – 200 km/s to stars in the thin disk. As the kick velocity the secondary receives is similar to the star's binary orbital velocity at the moment of the supernova explosion, we require the stars to be in tight binaries, with separations much smaller than the size of a pre–supernova red giant. This in turn implies that the binaries must have passed through a common envelope phase. Once ejected, such secondaries evolve into white dwarfs in a relatively short time if they are sufficiently massive.

Our evolutionary scenario is discussed in Section 2 of this paper and described in more quantitative detail in Appendix 1. In Section 3, we calculate the distribution of kicks received by the secondaries. In Section 4, we compute trajectories of secondaries within the Galaxy to determine whether they can produce a population of high–velocity objects similar to those seen by Oppenheimer et al. The likely production rate of such objects is discussed in Section 5 and Section 6 gives our conclusions.

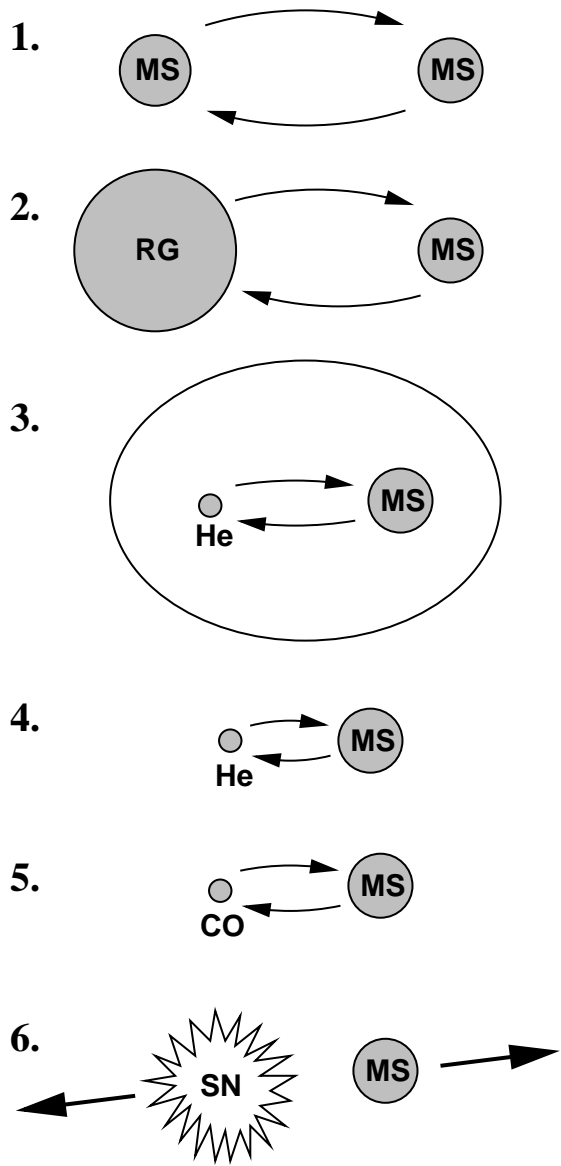

**Figure 1.** An evolutionary scenario for producing high-velocity main-sequence stars, some of which will later evolve into white dwarfs. Beginning with two main-sequence stars (phase 1), the primary evolves into a red giant (phase 2) filling its Roche lobe, leading to the onset of a common envelope phase (3) where the red-giant envelope engulfs the main-sequence star secondary and the helium-star core. The post-common-envelope system consists of a tight binary containing the helium-star core and the main-sequence star (phase 4). The compact star loses its helium envelope either via a wind or mass transfer, leaving a carbon-oxygen core (phase 5), which then explodes as a supernova, with the main-sequence star secondary often being ejected at high speed (phase 6).

## 2 EVOLUTIONARY SCENARIO

The evolutionary scenario for producing high-velocity stars considered in this paper is shown in Figure 1. We consider the evolution of binary stars where convective mass transfer occurs when the primary evolves off the main sequence and fills its Roche lobe. A common envelope phase follows. The secondary and the core of the primary spiral together as the envelope is ejected. After the common envelope phase, further mass transfer occurs if the primary (now a helium star) evolves to fill its Roche lobe for a second time. This (conservative) mass transfer increases the mass of the secondary but may also lead to the merger of the helium star and main-sequence star. Alternatively, if the primary fails to fill its Roche lobe, mass-loss occurs via a wind. In either case, the carbon-oxygen core of the primary remains. Providing the primary was initially sufficiently massive, it then explodes as a supernova. The secondary and the newly-formed neutron star become unbound in a large fraction of such binaries through the combined effects of mass-loss during the supernova explosion and the kick received by the neutron star on its formation. The kick velocity received by the secondary will be similar to its orbital velocity at the moment of the explosion. Hence in very wide binaries, where a common envelope phase is avoided, such velocities will be relatively low; too low to produce the high-velocity stars we require. However we will see in Section 4 that kick velocities $\gtrsim 100$ kms$^{-1}$ are realised in a subset of binaries which have passed through a common envelope phase. A large fraction of the high-velocity secondaries subsequently evolve into white dwarfs, especially if a large fraction of secondaries are relatively massive. High-velocity stars should also be visible before they evolve into white dwarfs. For example Hoogerwerf, de Bruijne & de Zeeuw (2001) and Magee et al (2001) both observe a population of high-velocity O and B-type main-sequence stars.

It is possible that the helium star in the wider systems does not have sufficient time to lose its entire He-rich envelope via a wind before exploding as a supernova. There would be two consequences of this. Firstly, the binary would be closer when the primary explodes as a supernova hence the secondary would have a higher orbital velocity. Secondly, the primary ejects more mass when it explodes, hence a larger fraction of binaries would be broken up.

There are a number of restrictions on the initial separations of binaries for the scenario described above. The system must initially be wide enough that the primary has evolved off the main sequence, and be convective, before it fills its Roche lobe. Yet the binary must not be too wide, otherwise no mass transfer will occur. We require that the main-sequence-star secondary not fill its Roche lobe at the end of the common envelope phase. If it does, we assume that the two stars merge and remove such binaries from further consideration.

The evolutionary scenario described above is discussed more quantitatively in Appendix 1.

## 3 POPULATION PRODUCED

In this section we consider the population of objects produced via the evolutionary scenario described in the previous section. If we assume that all stars of mass 8.3 M$_\odot \leq M_\star \leq$ 40 M$_\odot$ produce type II supernovae, and that the IMF is given by $dN/dM \propto M^{-2.35}$, the majority of supernovae originate from stars less massive than 20 M$_\odot$. For illustrative purposes, we consider binaries of primary mass $M_1 = 12$ M$_\odot$. In Figure 2 we plot the initial separations as a function of secondary mass, $M_2$, showing the various constraints for the evolution of binaries as described in Appendix 1. The lowest of the three solid lines represents the minimum separation required to avoid mass transfer whilst the primary is still on the main sequence. The top two solid lines are the limits represented by equation (A1). Systems above the uppermost of the three lines will remain detached throughout the evolution of the primary. Those below the uppermost line, but above the middle line, will undergo mass transfer after the primary has evolved off the main sequence and become convective (and thus enter a common envelope phase) while those below the middle line but above the bottom line will undergo mass transfer while the primary is still radiative. This will also lead to a common envelope phase if $M_2 \lesssim M_1/3$. In this case, the common envelope phase will result in the merger of the primary core and the secondary. The dotted line is the limit set by requiring that the main-sequence star secondary remains detached immediately after the common envelope phase (ie equation [A5], where to produce Figure 2 we have set $\alpha_{\rm CE}\lambda_{\rm CE} = 0.5$). The dashed line is the minimum initial separation to avoid the helium-star primary filling its Roche lobe when it evolves after the common envelope phase (ie equation [A6]). In the case of systems of primary mass, $M_1 = 12$ M$_\odot$, mass transfer from the Helium star leads to a merger with the main–sequence star secondary for $M_2 \lesssim 3$ M$_\odot$. Hence such systems will not produce high–velocity white dwarfs. The black dots in Figure 2 each represent systems evolving as described in the previous section, where the secondary is in a tight binary around the primary which explodes as a supernova. This region is limited from above by the requirement that the primary fills its Roche lobe (upper limit in equation [A1]) and from below by the requirement that the main-sequence star secondary remains detached immediately after the common envelope phase (ie equation [A5]). For systems where the post-CE binary is sufficiently wide that the helium star does not fill its Roche lobe, this star loses its envelope in a wind, with the binary separation increasing slightly as given by equation (A10). In addition to the constraints shown in Figure 2, there is a lower limit on the mass of the secondaries ($M_2 \geq 0.8 M_\odot$) if the secondaries are to evolve into white dwarfs during the age of the Galaxy. The dotted region in Figure 2 will shift slightly for other values of $\alpha_{\rm CE}\lambda_{\rm CE}$. However the range of binary separations at the moment of the supernovae explosion, and thus the distribution of kicks received by the secondary are similar.

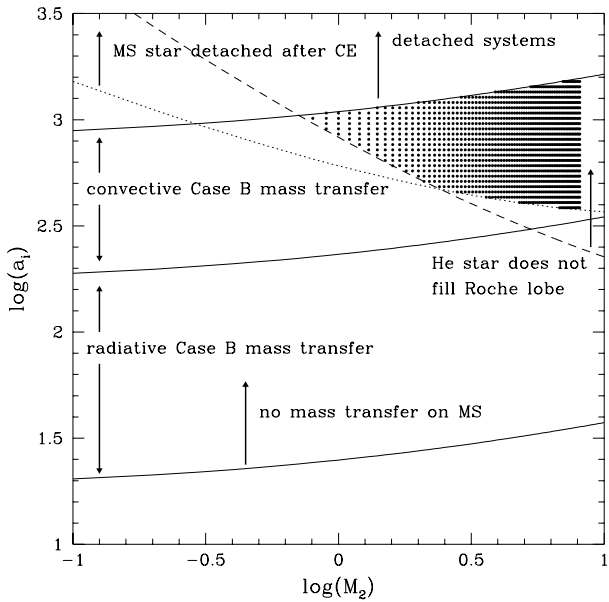

**Figure 2.** Plot of initial separation $a_i$, for $M_1 = 12$ M$_\odot$, as a function of secondary mass $M_2$(both in solar units) showing constraints for the formation of a binary as described in Section 3 and Appendix 1.

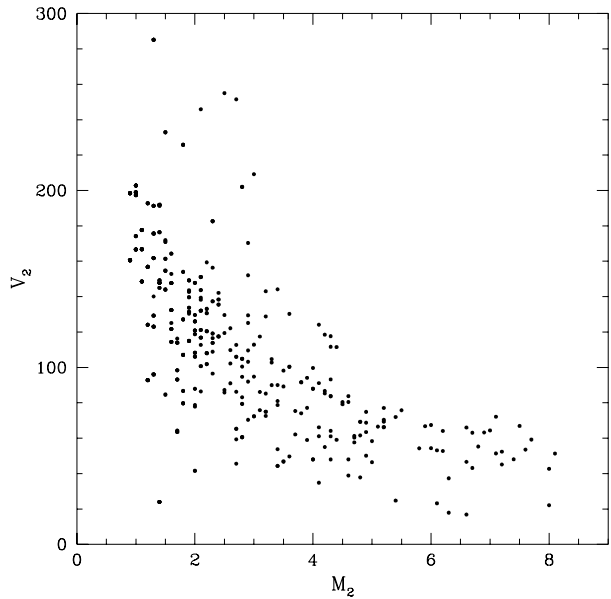

**Figure 3.** Plot of kick velocities received by secondaries (in km/s) contained in binaries which are broken up when the primary explodes as a supernova. The binary population is that shown in Figure 2 where the initial primary mass is 12 M$_\odot$ in all cases. The kick velocities are shown as a function of secondary mass $M_2$ (in solar units).

A similar picture to that shown in Figure 2 emerges for primary masses $M_1 \lesssim 12$ M$_\odot$. For more massive primaries, the helium-star primary never fills its Roche lobe in the systems of interest (as higher–mass stars produce higher–mass helium stars which evolve to have smaller radii).

We now consider the effects of the supernova explosion on the binaries. We allow for the effect of mass–loss from the primary (we assume the post-supernova mass of the primary is 1.4 M$_\odot$). We also include the kick received by the neutron star at formation, using a kick distribution from Hansen & Phinney (1997). Approximately two-thirds of the binaries identified in Figure 2 are broken up. We show the kick velocity received by the secondaries, as a function of secondary mass, in Figure 3. Here we have taken, as example, a Salpeter IMF for the secondary masses, ie $dN/dM_2 \propto M_2^{-2.35}$. We note from this figure that the kick velocity is often above 100 km/s. The distribution of kick velocities is similar for other values of primary mass. We also note that in all cases $M_2 \gtrsim 0.9$ M$_\odot$, even though here we have assumed a Salpeter–like distribution for the secondary masses. The median mass for the secondaries is $\sim 3$ M$_\odot$ hence we would expect the vast majority of ejected secondaries to have evolved into white dwarfs by today. In Section 4 we use this kick distribution to compute the resulting spatial distribution of secondaries throughout the Galaxy.

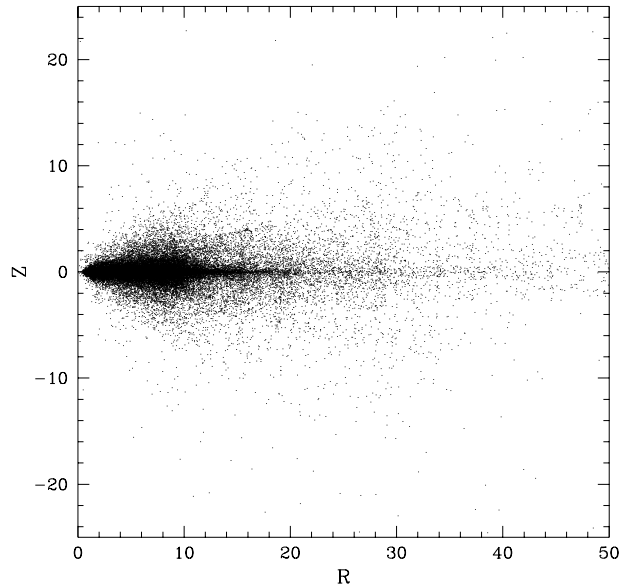

**Figure 4.** The positions of 1000 secondaries which have received kicks drawn randomly from the distribution shown in Figure 3 ($M_1 = 12$ M$_\odot$ and initial secondary masses are drawn from a Salpeter distribution). The trajectories have been followed for 10 Gyr and the positions noted every 50 Myr. The positions are shown in cylindrical coordinates, with units of kpc.

## 4  TRAJECTORIES OF HIGH–VELOCITY WHITE DWARFS

Using the model for the Galactic potential described in Appendix 2, we now compute the subsequent trajectories of the secondaries of the binaries which are broken up when the primary explodes as a supernova. For illustration we again consider binaries of initial primary mass $M_1 = 12$ M$_\odot$, with the secondaries having a Salpeter–like distribution of masses. Hence we use the kick distribution given in Figure 3. We will assume that the initial (circular) orbital radius of the binary in the Galaxy is drawn uniformly from the range 1 kpc – 10 kpc. We then apply to the secondary a kick drawn randomly from our kick distribution (and directed in a random direction with respect to the original motion within the Galaxy) and integrate its motion within the Galactic potential for 10 Gyr, writing out its position and velocity every 50 Myr. We consider 1000 trajectories. A plot of all the positions (in cylindrical coordinates) is shown in Figure 4. We note that a large fraction of positions are relatively close to their birth site. In Figure 5 we plot the tangential and radial velocities of all stars within the limits 8 kpc $\leq R \leq$ 9 kpc and $-0.5$ kpc $\leq z \leq 0.5$ kpc. We see that the distribution of velocities in the $U - V$ plane is encouragingly similar to that reported by Oppenheimer et al. The distribution shown in Figure 5 will clearly depend on the distribution of kicks received by the secondaries, which will depend on the distribution of secondary masses as shown in Figure 3.

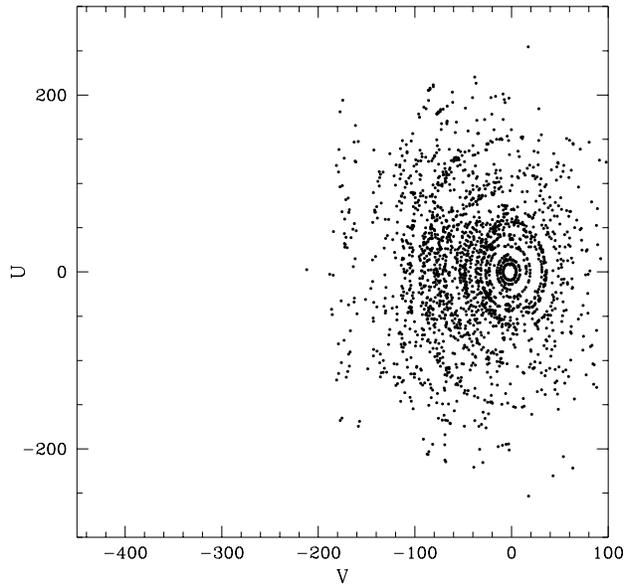

**Figure 5.** Plot of velocities (in kms$^{-1}$) in the U–V plane for a subset of positions in Figure 4 given by 8.0 kpc $\leq R \leq$ 9.0 kpc and $-0.5$ kpc $\leq z \leq 0.5$ kpc.

## 5  PRODUCTION RATE

In the previous section we showed that a reasonable set of assumptions for the kick distribution and the initial positions of the binaries yields a distribution of stellar velocities in the solar neighbourhood not unlike that seen by Oppenheimer et al. However, we have not yet addressed the issue of formation rates of such systems. The formation rate is dependent on a number of uncertain factors: the global supernova rate within the Galaxy, the distribution of primary masses, the binary fraction, the distribution of secondary masses for a given primary mass, the distribution of initial binary separations, and the initial spatial distribution of binaries within the Galaxy.

Given all these uncertainties, it is impossible to give a precise figure for the local number density of white dwarfs produced as described in this paper. Rather we can obtain a reasonable idea of the order of magnitude of the likely population given a reasonable set of assumptions. Experimentally we note that some 0.3% of systems plotted in Figure 4 are within the limits 8.4 kpc $\leq R \leq$ 8.6 kpc and $-0.1$ kpc $\leq z \leq 0.1$ kpc. The volume of this region is $V \sim 2 \times 10^9$ pc$^3$. If we assume a (slightly high) supernova rate of $10^{-1}$/yr, the total number of type II supernovae over the entire life of the Galaxy is $N_{\rm SN} \sim 1.5 \times 10^9$. The local number density of ejected secondaries is given by $N_{\rm SN}/V \sim 2 \times 10^{-3} f_2$ pc$^{-3}$, where $f_2$ is the fraction of type II supernovae occurring in the sorts of binaries discussed in Section 3. In other words the process described here will produce the population of objects observed by Oppenheimer et al if $f_2 \gtrsim 0.05$. It is quite possible that another set of assumptions would lead to an even higher production rate than the one derived here.

## 6 CONCLUSIONS

We have described a process producing a population of high–velocity stars from binaries in the thin disk. If the primary fills its Roche lobe after leaving the main sequence, convective mass transfer leads to a common envelope phase. The separation between the core of the primary and the secondary decreases significantly as the envelope is ejected. A second phase of mass transfer may occur if the primary (now a helium star) evolves to fill its Roche lobe. This phase is conservative and therefore increases the mass of the secondary. Alternatively if the helium star fails to fill its Roche lobe, mass loss will occur via a wind. In both cases, the carbon-oxygen core of the primary remains.

For initial primary masses, $M_{1,i} \geq 8.3$ M$_\odot$, this carbon-oxygen core then explodes as a type II supernova. Most of the binaries become unbound through the combined effects of mass-loss in the supernova explosion and the kick received by the neutron star at formation.

We have computed the distribution of kick velocities for these secondaries. Considering a population of binaries initially on circular orbits within the disk of the Galaxy, we apply kick velocities drawn from the distribution in random directions and integrate the motion of the ejected secondaries within the Galaxy. We show that for a reasonable set of assumptions, we are able to obtain a population of high-velocity stars similar to that seen by Oppenheimer et al. There are number of uncertainties associated with calculating a production rate of these stars, but for a reasonable set of assumptions, we show that this population may make a significant contribution to the population observed by Oppenheimer et al.

## ACKNOWLEDGMENTS
MBD gratefully acknowledges the support of a URF from the Royal Society. Theoretical astrophysics research at Leicester is supported by a rolling grant from the UK Particle Physics & Astronomy Research Council.

## APPENDIX 1: EVOLUTIONARY SCENARIO

In order for the late (convective) Case B mass transfer from the primary to occur, leading to the onset of a common envelope phase, we require that the primary fill its Roche lobe, but not until it has a radius larger than the maximum allowed for early (radiative) Case B mass transfer. This requires

$$\frac{R_{\rm max,rB}(M_{1,i})}{f_1(q_i)} < a_i < \frac{R_{\rm max,B}(M_{1,i})}{f_1(q_i)} \tag{A1}$$

where $q_i = M_{1,i}/M_{2,i}$ and $R_{\rm max,B}(M_{1,i})$ is the maximum radius allowed for Case B mass transfer for a star of mass $M_{1,i}$. The radius of star 1 when it fills its Roche lobe is given by $R_{1,i} = f_1(q_i)a_i$, where $f_1(q) = 0.49q^{2/3}/(0.6q^{2/3} + \ln(1 + q^{1/3}))$ (Eggleton 1983).

The mass of the primary after the common envelope phase is given by

$$M_{1,\rm CE} = M_{\rm He}(M_{1,i}) = aM_{1,i}^b \tag{A2}$$

where $a = 0.125$ and $b = 1.4$ (van den Heuvel 1994).

The inspiral during the common envelope phase may be computed by equating the change in total energy of the two stars to the binding energy of the envelope up to an efficiency $\alpha_{\rm CE}$, ie $E_{\rm env} = \alpha_{\rm CE}\Delta E$. Here $E_{\rm env}$ can be written in the following form (Webbink 1984)

$$E_{\rm env} = \frac{GM_{1,i}(M_{1,i} - M_{1,\rm CE})}{\lambda_{\rm CE} f_1(q_i) a_i} \tag{A3}$$

Combining equation (A2) with the expression for $\Delta E$, after some rearrangement we arrive at the following expression for the inspiral during the common envelope phase

$$F_{i\to\rm CE} = \frac{a_{\rm CE}}{a_i} = \left(\frac{2M_{1,i}(M_{1,i} - M_{1,\rm CE})}{\alpha_{\rm CE}\lambda_{\rm CE} f_2(q_i) M_{1,\rm CE} M_{2,i}} + \frac{M_{1,i}}{M_{1,\rm CE}}\right)^{-1} \tag{A4}$$

We require that the main-sequence-star secondary not fill its Roche lobe at the end of the common envelope phase. If it does, we assume that the two stars merge and remove such binaries from further consideration.

$$a_i > \frac{R_{\rm MS}(M_{2,\rm CE})}{F_{i\to\rm CE} f_2(q_{\rm CE})} \tag{A5}$$

where $R_{\rm MS}(M_{2,\rm CE})$ is the radius of the main-sequence-star secondary and $q_{\rm CE} = M_{1,\rm CE}/M_{2,\rm CE}$.

The primary will now be a helium star. Such stars will expand as they evolve and may, in turn, fill their Roche lobe if

$$a_{\rm CE} < \frac{R_{\rm max,He}(M_{1,\rm CE})}{f_1(q_{\rm CE})} \tag{A6}$$

Under these conditions, conservative (Case BB) mass transfer will follow, where the envelope of the helium star will be transferred to the secondary leaving the carbon-oxygen core of the primary. The masses of the two stars $M_{1,\text{CO}}$ and $M_{2,\text{PSN}}$ after such a phase of mass transfer are given by

$$M_{1,\text{CO}} = M_{\text{CO}}(M_{1,\text{He}}) \tag{A7}$$

$$M_{2,\text{PSN}} = M_{2,\text{i}} + (M_{1,\text{He}} - M_{1,\text{CO}}) \tag{A8}$$

The ratio of the separation after this phase to the separation after the common envelope phase is given by

$$F_{\text{CE}\to\text{PSN}} = \frac{a_{\text{PSN}}}{a_{\text{CE}}} = \frac{M_{1,\text{CE}}^2 M_{2,\text{i}}^2}{M_{1,\text{CO}}^2 M_{2,\text{PSN}}^2} \tag{A9}$$

If, on the other hand, the He star never fills its Roche lobe, the envelope will be lost via a wind. Leaving behind the carbon-oxygen core, while the secondary mass will be unchanged (ie $M_{2,\text{PSN}} = M_{2,\text{CE}}$). The ratio of the separation after this phase to the separation after the common envelope phase is given by

$$F_{\text{CE}\to\text{PSN}} = \frac{a_{\text{PSN}}}{a_{\text{CE}}} = \frac{M_{1,\text{He}} + M_{2,\text{CE}}}{M_{1,\text{CO}} + M_{2,\text{PSN}}} \tag{A10}$$

The functions $R_{\max,\text{rB}}(M)$, $R_{\max,\text{B}}(M)$, $R_{\max,\text{He}}(M)$, and $M_{\text{CO}}(M)$, were tabulated from Habets (1986), Paczynski (1971) and Bressan et al (1993) and linear interpolation was used to evaluate them for a particular mass, $M$.

In order for the primary to explode as a Type II supernova, we require that the core mass exceeds the Chandrasekhar limit, ie $M_{1,\text{CO}} > M_{\text{CH}}$. For the assumptions here (equations [A2] and [A7]), this is equivalent to $M_{1,\text{i}} > 8.3$ $M_\odot$. The mass-loss (from the supernova explosion) and the kick received by the neutron star may unbind the binary.

If the binary is broken up, the kick velocity of the secondary will be similar to its orbital velocity at the moment of the supernova explosion, which is given by

$$V_{2,\text{PSN}} = \frac{M_{1,\text{CO}}}{M_{1,\text{CO}} + M_{2,\text{PSN}}} \sqrt{\frac{G(M_{1,\text{CO}} + M_{2,\text{PSN}})}{a_{\text{PSN}}}} \tag{A11}$$

**APPENDIX 2: THE GALACTIC POTENTIAL**

The Galactic potential can be modelled as the sum of three potentials. The spheroid and disc components are given by

$$\Phi_s(R, z) = \frac{GM_s}{(R^2 + [a_s + (z^2 + b_s^2)^{1/2}]^2)^{1/2}} \tag{A12}$$

$$\Phi_d(R, z) = \frac{GM_d}{(R^2 + [a_d + (z^2 + b_d^2)^{1/2}]^2)^{1/2}} \tag{A13}$$

where $R^2 = x^2 + y^2$. The component from the Galactic halo can be derived assuming a halo density distribution, $\rho_h$, given by

$$\rho_h = \frac{\rho_c}{1 + (r/r_c)^2} \tag{A14}$$

where $r^2 = x^2 + y^2 + z^2$. The above density distribution yields the potential

$$\Phi_h = -\frac{GM_c}{r_c}\left[\frac{1}{2}\ln\left(1 + \frac{r^2}{r_c^2}\right) + \frac{r_c}{r}\operatorname{atan}\left(\frac{r}{r_c}\right)\right] \tag{A15}$$

where $M_c = 4\pi\rho_c r_c^3$. The total Galactic potential is the sum

$$\Phi = \Phi_s + \Phi_d + \Phi_h \tag{A16}$$

Following Paczynski (1990), we use the following choice of parameters:

$$a_s = 0, \ b_s = 0.277 \text{ kpc}, \ M_s = 1.12 \times 10^{10} M_\odot, \tag{A17}$$

$$a_d = 3.7 \text{ kpc}, \ b_d = 0.20 \text{ kpc}, \ M_d = 8.07 \times 10^{10} M_\odot, \tag{A18}$$

$$r_c = 6.0 \text{ kpc}, \ M_c = 5.0 \times 10^{10} M_\odot, \tag{A19}$$

Because of the cylindrical symmetry of the potential, the integration of the trajectories can be simplified to consider the evolution of the $z$ and $R$ components only, as given below

$$\begin{aligned}\frac{dR}{dt} &= v_R, \quad \frac{dz}{dt} = v_z, \\ \frac{dv_R}{dt} &= \left(\frac{\partial \Phi}{\partial R}\right)_z + \frac{j_z^2}{R^3}, \quad \frac{dv_z}{dt} = \left(\frac{\partial \Phi}{\partial z}\right)_R\end{aligned} \tag{A20}$$

where the $z$ component of the angular momentum, $j_z = Rv_\phi$.